\documentclass[12pt]{article}
\usepackage{amssymb,latexsym,epsfig,cite}

\newcommand{\be}{\begin{equation}}
\newcommand{\ee}{\end{equation}}

\begin{document}

\thispagestyle{empty}

\baselineskip=15pt

\begin{center}
{\Large \bf 
Non--hermitian susy hydrogen--like\\[1ex]
Hamiltonians with real spectra}
\end{center}

\vskip1cm

\begin{center}
{Oscar Rosas-Ortiz\footnote{Corresponding Author. E--mail:
{\tt orosas@fis.cinvestav.mx}} and Rodrigo
Mu\~noz\footnote{E--mail: {\tt rodrigom@fis.cinvestav.mx}}}\\[1ex]
{\footnotesize
Departamento de F\'{\i}sica, CINVESTAV-IPN, A.P. 14-740,
07000 M\'exico D.F., Mexico}
\end{center}

\begin{center}
\begin{minipage}{14cm}
{\footnotesize
{\bf Abstract.} It is shown that the radial part of the hydrogen
Hamiltonian factorizes as the product of two not mutually adjoint first
order differential operators plus a complex constant $\epsilon$. The
1--susy approach is used to construct non-hermitian operators with
hydrogen spectra. Other non--hermitian Hamiltonians are shown to admit an
extra `complex energy' at $\epsilon$. New self--adjoint hydrogen-like
Hamiltonians are also derived by using a 2--susy transformation with
complex conjugate pairs $\epsilon$, $\bar \epsilon$.}
\end{minipage}
\end{center}

\vskip0.5cm

\noindent 
{\footnotesize
[1ex]
}

%%%%%%%%%%%%%%%%%%%%%%%%%%%%%%%%%%
\section{Introduction}

The supersymmetric Quantum Mechanics (SUSY QM) has grown on the
factorization and intertwining methods \cite{Car79} applied to transform
the physical Hamiltonians. It yields new exactly solvable potentials which
are either strictly isospectral to the initial one because of broken SUSY,
or almost isospectral due to unbroken SUSY (see the recent reviews
\cite{Lim02}.) The higher order SUSY QM amended the conviction that the
excited states can not be used to generate non--singular susy partners
\cite{And93, Fer97, Fer98, Ros98}. Some applications deal with singular
\cite{Mar98}, soliton--type \cite{Mie00}, periodic \cite{Fer00} and other
potentials \cite{Aiz97}. Of special interest is the confluent algorithm
\cite{Mie00, Fer02b} for which the second order procedure is applied to
add a single level at an arbitrary point of the energy axis. However,
almost all the works on the subject make use of transformation functions
with the real factorization constants $\epsilon$ and with the
factor--operators being always mutually adjoint.

The case of complex $\epsilon$ has not been studied to the desired extent.
Exceptional cases seem \cite{Bag95, Fer02, Bay96}, where the 2--susy
treatment with $\epsilon \in {\mathbb C}$ is formulated to obtain either
hermitian or non--hermitian susy partners of a given initial Hamiltonian.
Indeed, one of our purposes is to show that the Hamiltonians $H$ can be
factorized as the product of two not mutually adjoint first order
operators $A$, $B$, plus a complex constant even for self--adjoint $H$.
The method is not limited to the Hamiltonians possessing the lower
spectral bound; it can also yield non--hermitian Hamiltonians with complex
potentials.  Although the typical Hamiltonians of QM are hermitian,
non--hermitian ones appear in Molecular Physics and Quantum Chemistry
\cite{Mal85}, Superconductivity \cite{Hat96}, Quantum Field Theory
\cite{Ben99a} and other domains \cite{Hoe79}. The fact that they admit
real eigenvalues for which the associated eigenfunctions are
square--integrable \cite{Ben98} has been the basis of recent studies on
{\it PT\/}--symmetry \cite{Ben99b, Dor01}, pseudo--hermiticity
\cite{Mos02, Ahm02} and diverse physical models \cite{Ket01}.

In this paper we illustrate these facts by constructing the hermitian and
non--hermitian susy partners of the radial part ($H_{\ell}$) of the
hydrogen Hamiltonian. Even though our non--hermitian operators are not
{\it PT\/}--invariant, we shall see that a class of them has real
eigenvalues identical to the hydrogen energies. The reality of the
spectrum in this case is due to the breaking of supersymmetry.

It will be shown that there is another class of non--hermitian operators
having an extra square--integrable eigenfunction associated with $\epsilon
\in {\mathbb C}$. In this case, the `complex energy' $\epsilon$ arises
from the unbroken supersymmetry and, up to now, does not have a well
established physical meaning (but see \cite{last}). Unlike the
phase--equivalent complex potentials \cite{Bay96}, the new `bound state'
associated with $\epsilon$ is nodeless. Moreover, in counterdistinction
with the formalism of {\it PT\/}--symmetry and pseudo--hermiticity, where
complex energies appear in conjugate pairs, it turns out that $\bar
\epsilon$ does not belong to the spectrum of the susy partner of
$H_{\ell}$ generated through $\epsilon$.

In general, we shall see that the susy transformation is adequate to
analytically determine normalizable eigenfunctions of non--hermitian
Hamiltonians, including those with the complex energies. In this sense,
our `complex susy transformation' seems an analytical complement of the
numerical techniques previously reported (see \cite{Bay96} and references
therein.) The eigenfunction connected with $\epsilon$ is then removed by
iterating the procedure in order to construct hermitian 2--susy partners
of $H_{\ell}$.

The paper is organized as follows: Section~2 introduces the atypical
factorizations $H_{\ell} = AB + \epsilon$, where $\epsilon \in {\mathbb
C}$ and the first order differential operators $A$ and $B$ are not
mutually adjoint. Sections 3 and 4 are devoted to the construction of
non--hermitian 1--susy and hermitian 2--susy partners, $\mathsf{H}(\zeta)$
and $\widetilde H$ respectively, of $H_{\ell}$. Final remarks and
discussion are given in Section~5.

%%%%%%%%%%%%%%%%%%%%%%%%%%%%%%%%%%
\section{The complex-type factorization method}

Let us consider a single electron in the field produced by a nucleus with
$Z$ protons. We shall use ${\cal E} = Z/2r_B$ and $r_B = \hbar^2/Ze^2m$
for the units of energy and coordinates respectively. The corresponding
time--independent Schr\"odinger equation reduces to $H_{\ell} \, \psi_{n,
\ell}(r) = E_n \, \psi_{n, \ell}(r)$, with solutions
\be
\psi_{n, \ell}(r) = C_{n, \ell} \, r^{\ell +1} \, e^{-r/n} \, {}_1F_1(\ell
+1 -n, 2 \ell + 2; 2r/n), \qquad E_n = -1/n^2
\label{solfis}
\ee
where ${\mathbb N} \ni n=\ell + s+1$; $\ell=0,1,2,...,n-1$; $s \in
{\mathbb Z}^+$; $C_{n, \ell}$ is the normalization constant,
${}_1F_1(a,c,z)$ is the Kummer's function and $L^2({\mathbb R}^+, 4\pi)
\ni \psi_{n, \ell}(r) \equiv rR_{n, \ell}(r)$, with an inner product
defined by $\langle\psi, \phi \rangle = 4 \pi \, \int_0^{+\infty} \bar
\psi(r) \phi(r) dr < \infty$ and boundary conditions at $r=0$:
$\psi(0)=0$, $\psi'(r)=R(0)$. The effective potential $V_{\ell}(r)$ has
the domain ${\cal D}_V =[0,\infty)$ and
\be
H_{\ell} \equiv -\frac{d^2}{dr^2} + V_{\ell}(r) = -\frac{d^2}{dr^2} +
\frac{\ell(\ell +1)}{r^2} -\frac{2}{r}.
\label{ham1}
\ee

The Hamiltonian (\ref{ham1}) is factorized as follows:
\be
H_{\ell} = AB + \epsilon
\label{factor1}
\ee
where the factorization constant is a complex number ${\mathbb C} \ni
\epsilon := \epsilon_1 + i \epsilon_2$; $\epsilon_1, \epsilon_2 \neq 0\in
{\mathbb R}$ and the first order operators $A, B$, are not mutually
adjoint (compare \cite{Ros98, Fer84}):
\be
A:=- \frac{d}{dr} + \beta(r), \qquad B:= \frac{d}{dr} + \beta(r)
\label{opera1}
\ee
with $\beta$ a complex--valued function fulfilling
\be
-\beta'(r) + \beta^2(r) + \epsilon = V_{\ell}(r).
\label{riccati1}
\ee
The Riccati equation (\ref{riccati1}) is solved by means of the
logarithmic transformation $\beta(r) = - \frac{d}{dr} \ln u(r)$ for which
$u$ is the most general eigenfunction of $H_{\ell}$ (not necessarily
normalizable) belonging to $\epsilon \equiv -k^2$;  ${\mathbb C} \ni k=k_1
+ i k_2$;  $k_1, k_2 \in {\mathbb R}$:
\be
\begin{array}{ll}
u(r) = & r^{\ell +1} e^{-kr}\, f(r);\\[1ex]
f(r) := & \alpha \,{}_1F_1(\ell +1 -1/k, 2\ell +2, 2kr) +
\zeta \, U(\ell +1 -1/k, 2\ell +2, 2kr)
\end{array}
\label{u}
\ee
where $\alpha$ and $\zeta$ are complex constants and $U(a,c,z)$ is the
logarithmic hypergeometric function. The global behaviour of these
$u$-functions is analized in Appendix.

Hence, for the $\beta$--function we have:
\be
\beta(r)= -\frac{\ell +1}{r} + k + \Omega(r); \qquad \Omega(r):=-
\frac{d}{dr} \, \ln f(r).
\label{beta}
\ee
A convenient expression for $\beta_1(r)$, $\beta_2(r)$, the real and
imaginary parts of $\beta(r)$ respectively, can be found in Appendix.

%%%%%%%%%%%%%%%%%%%%%%%%%%%%%%%%%%
\section{New complex hydrogen-like potentials}

Let us consider the value $\epsilon$ fixed. By convenience we shall
make explicit the dependence of $\Omega$ on $\zeta$. Now, let us reverse
the order of the factors in (\ref{factor1}):
\be
BA + \epsilon = -\frac{d^2}{dr^2} + V_{\ell + 1}(r) + 2 \Omega'(r; \zeta)
\equiv -\frac{d^2}{dr^2} + \mathsf{V}(r;\zeta):= \mathsf{H}(\zeta)
\label{factor2}
\ee
where $\mathsf{H}(\zeta)$ is a non--hermitian second order differential
operator and we have used (\ref{riccati1}) and (\ref{beta}). The next step
is to solve the related eigenvalue equation:
\be
\mathsf{H}(\zeta) \, \Psi = \lambda \Psi, \qquad
\lambda = \lambda_1 + i \, \lambda_2, \qquad
\lambda_1, \lambda_2 \in {\mathbb R}.
\label{scro2}
\ee
The dependence of $\Psi$ and $\lambda$ on $\zeta$ will be dropped for
simplicity. Notice that equations (\ref{factor1}) and (\ref{factor2})
imply an intertwining between the Hamiltonian $H_{\ell}$ and
the non--hermitian operator $\mathsf{H} (\zeta)$:
\be
\mathsf{H}(\zeta) \, B = B H_{\ell}, \qquad H_{\ell} A = A
\mathsf{H}(\zeta)
\label{intertwin1}
\ee
Thereby, one sees that $\Psi \propto B \varphi$ is a solution of
(\ref{scro2}) if $\varphi$ satisfies $H_{\ell} \, \varphi = \lambda
\varphi$, while $A$ reverses the action of $B$. Now, the general form of
$\varphi$ is obtained by taking $\epsilon = -k^2$ for $\lambda =
-\kappa^2$, and $u(r)$ for $\varphi(r)$ in (\ref{u}). Hence, we have:
\be
\varphi(r) = r^{\ell +1} e^{-\kappa r} \{ C \,{}_1F_1(\ell +1 -1/\kappa,
2\ell +2, 2\kappa r) + D\, U(\ell +1 -1/\kappa, 2\ell +2, 2\kappa r)\}
\label{varphi1}
\ee
with $C$ and $D$ arbitrary complex constants. Therefore
\be
\Psi \propto B \varphi = \frac{W(u, \varphi)}{u}.
\label{Psi1}
\ee
We are looking now for the constraints on $\alpha, \zeta, \lambda, C$ and
$D$ leading to square--integrable $\Psi$. First, consider $\lambda \neq
\epsilon$ ({\it i.e.\/}, $\kappa \neq k$); the behaviour of $\Psi$ near
the origin is
\be
\Psi(r \sim 0) \propto \left\{
\begin{array}{ll}
- D \, \frac{(2 \ell +1)}{(2 \kappa)^{2 \ell +1}} \, \frac{\Gamma(2 \ell
+1)}{\Gamma(\ell +1 -1/\kappa)} \, \frac{1}{r^{\ell +1}}, \qquad & \zeta
=0, \, \, \alpha \neq 0 \\[2ex]
D \, \frac{(k -\kappa)}{(2 \kappa)^{2 \ell +1}} \, \frac{\Gamma(2 \ell
+1)}{\Gamma(\ell +1 -1/\kappa)} \, \frac{1}{r^{\ell}} & \zeta \neq 0, 
\, {\rm arbitrary} \,\, \alpha
\end{array}
\right.
\label{psi}
\ee
Thus, $\Psi$ becomes divergent at $r=0$ except if either (I) $\lambda =
-\kappa^2$ is real and $\kappa^{-1} = \ell +s +1$, $s \in {\mathbb Z}^+$,
or (II) $\lambda$ is complex but $D=0$. Let us pay some attention to these
conditions

\noindent
{\bf CASE I (real $\lambda$):} Let us fix ${\mathbb R} \ni
\kappa^{-1}=\ell +s +1=n$, $n \in {\mathbb N}$. In this case, in
(\ref{varphi1}) ${}_1F_1(a,c;z)$ and $U(a,c;z)$ are essentially the same
function. Therefore, one can take $D=0$ and $C=C_{n, \ell}$ (see equation
(\ref{solfis})), so $\varphi (r) = \psi_{n, \ell}(r)$ and the functions
(\ref{Psi1}) behave asymptotically as $\Psi(r) \propto \psi_{n, \ell} \,
(r)_{r \rightarrow \infty}$. On the other hand, a straightforward
calculation shows that these functions obey the following boundary
conditions at $r=0$: $\Psi(0, \zeta) = R(0)$; $\Psi'(0, \zeta=0) = -
\delta_{\ell \, 0}$; $\Psi'(0, \zeta \neq 0) = - \ell (\frac{\delta_{\ell
\, 0}}{r} + \delta_{\ell \, 1})$, with $\delta_{\ell \, n}$ the kronecker
delta. Thus
\be
\Psi(r; \zeta) \propto \left[ k - \frac{1}{n} + \frac{d}{dr} \ln \,
\left( \frac{{}_1F_1(\ell + 1 - n, 2 \ell +2; 2r/n)}{f(r)} \right) \right]
\psi_{n,\ell}(r)
\label{Psi2}
\ee
are square--integrable eigenfunctions of $\mathsf{H} (\zeta)$ with the
real eigenvalues $\lambda = -1/n^2 = E_n$. 

\noindent
{\bf CASE II (complex $\lambda$):} For $\kappa \in {\mathbb C}$ and $D=0$
the function (\ref{Psi1}) behaves asymptotically as follows:
\be
\Psi(r) \sim \left\{
\begin{array}{ll}
C \, \frac{(\kappa \mp k)}{(2 \kappa)^{\ell +1 + 1/\kappa}} \,
\frac{\Gamma(2 \ell + 2)}{\Gamma(\ell +1 -1/\kappa)} \, \frac{e^{\kappa
r}}{r^{1/\kappa}} \quad & \kappa_1 >0, \, 
\left\{
\begin{array}{l}
k_1 >0\\
k_1<0
\end{array}
\right.
\\[5ex]
\mp C \, \frac{(k \pm \kappa)}{(-2 \kappa)^{\ell +1 - 1/\kappa}} \,
\frac{\Gamma(2 \ell + 2)}{\Gamma(\ell +1 + 1/\kappa)} \, \frac{e^{-\kappa
r}}{r^{-1/\kappa}} \qquad & \kappa_1<0, \,
\left\{
\begin{array}{l}
k_1 >0\\
k_1<0
\end{array}
\right.
\end{array}
\right.
\label{Psi3}
\ee
which always diverges for $r \rightarrow \infty$. Hence, there is no
function $\Psi \in L^2({\mathbb R}^+, 4 \pi)$ solving (\ref{scro2}) for
a complex $\lambda \neq \epsilon$ ({\it i.e.\/} $\kappa \neq k$) and
$D=0$.

The formula (\ref{Psi2}) therefore gives all square integrable solutions
$\{ \Psi(r;\zeta) \}$ of (\ref{scro2}) for $\lambda \neq \epsilon$.
Concerning the case $\lambda = \epsilon$, we see from equation
(\ref{factor2}) that any $\Psi_{\epsilon}(r)$ in the one--dimensional
kernel of $A$ is an eigenfunction of $\mathsf{H}(\zeta)$ belonging to
$\epsilon$. After a simple calculation one gets
\be
\Psi_{\epsilon}(r) \propto \frac{1}{u(r)}.
\label{missing}
\ee
This function can be in $L^2({\mathbb R}^+, 4\pi)$ for appropriate values
of $\alpha$ and $\zeta$. In such a case, $\Psi_{\epsilon}$ is an extra
square-integrable eigenfunction of ${\mathsf H} (\zeta)$ associated with
$\epsilon \in {\mathbb C}$. Finally, a straighforward calculation shows
that, although $\{ \Psi(r;\zeta), \Psi_{\epsilon}(r) \}$ are elements of
$L^2({\mathbb R}^+, 4\pi)$, they do not form an orthogonal set. A
discussion on this kind of properties of the inner product is given in
\cite{Ket01}. The next subsections analize these conditions and classify
the resulting potentials according to their spectra.

%%%%%%%%%%%%%%%%%%%%%%%%%%%%%%%%%%%
\subsection{The real spectrum}

From Figure~1a, one sees that the behaviour of the complex function
$\mathsf{V}(r; \zeta)$, for $\alpha \neq 0$ and $\zeta = 0$, is given by
\[
\mathsf{V}(r;\zeta=0) \sim \left\{
\begin{array}{ll}
V_{\ell +1}(r), \qquad & \mbox{\rm for} \, \, r \sim 0,\\[1ex]
0, \qquad & r \rightarrow \infty.
\end{array}
\right.
\]
Consistently with the case ({\bf A}) of Appendix, the function
$\Psi_{\epsilon}(r)$ diverges at $r=0$. Hence, $\Psi_{\epsilon}(r)$ is not
in $L^2({\mathbb R}^+, 4\pi)$ and the discrete spectrum $\sigma_d({\mathsf
H}_{\ell+1})$ of ${\mathsf H}(\zeta = 0)$ is exactly the same as that of
the hydrogen atom $\sigma_d({\mathsf H}_{\ell+1}) = \sigma_d(H_{\ell})$.
Figure~1b depicts the behaviour of $\psi_{n, \ell}$ and the related
function (\ref{Psi2}) for one of the excited states.

\medskip
%%%%%%%%%%%%%%%%%%%%%%%%%
\begin{figure}[ht]
\centering 
\hskip-0.5cm \epsfig{file=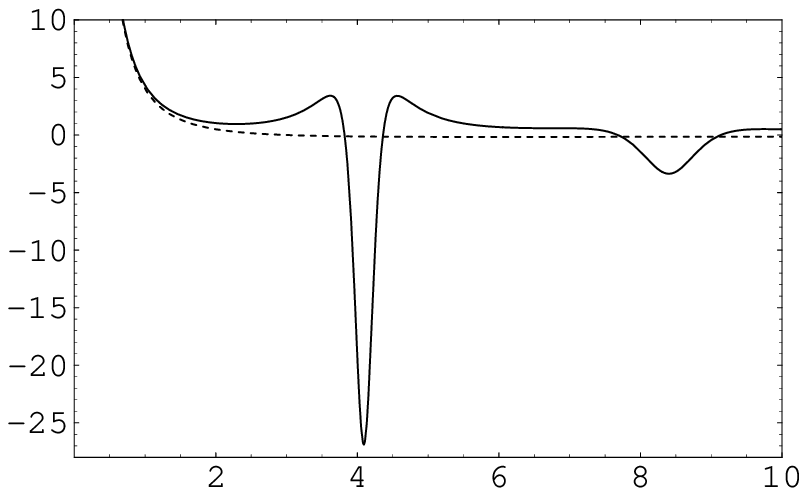, height=4cm}
\hskip1cm \epsfig{file=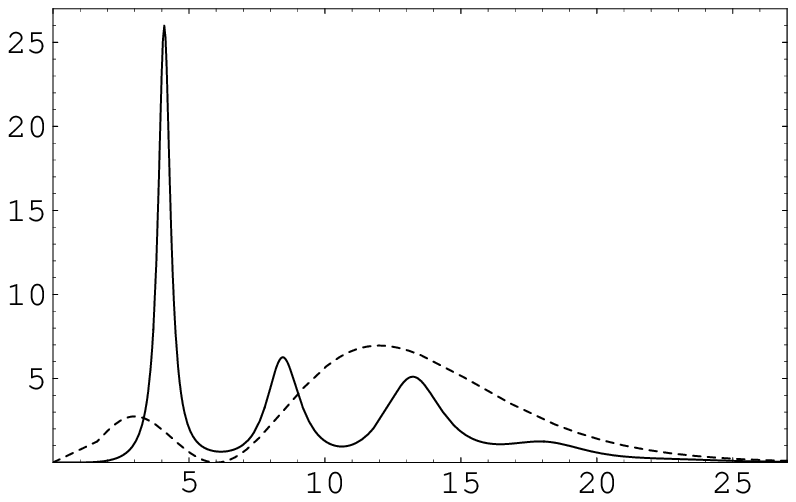, height=4cm}
\centerline{(a) \hskip6.7cm (b)}
\caption{\footnotesize {\bf (a)} The potential $V_{\ell + 1}(r)$ (dashed
curve) and the real part of $\mathsf{V}(r;\zeta)$ for $\ell =1$,
$\epsilon=-(0.1 + i \, 0.5)^2$, $\alpha=1$, and $\zeta =0$. {\bf (b)} The
corresponding unnormalized $\vert \psi_{3,1} (r) \vert^2$ (dashed curve)
and its susy partner $\vert \Psi(r;\zeta=0) \vert^2$ for $E_3=-1/9$ and
the same values of the parameters.}
\end{figure} 
%%%%%%%%%%%%%%%%%%%%%%%%%
\medskip

\noindent
Another class of complex potentials ${\mathsf V}(r; \zeta)$ sharing the
same spectrum as $H_{\ell}$ is obtained by considering the case ({\bf B})
of Appendix.

%%%%%%%%%%%%%%%%%%%%%%%%%%%%%%%%%%%%%%%%%%%%%%%%%
\subsection{Complex potentials admitting complex `energies'}

Let us consider $\zeta \neq 0$. Then the complex function $\mathsf{V}(r;
\zeta)$ behaves as shown on Figure~2a, {\it i.e.\/}:
\[
\mathsf{V}(r;\zeta \neq 0) \sim \left\{
\begin{array}{ll}
V_{\ell -1}(r), \qquad & \mbox{\rm for} \, \, r \sim 0, \, \,
\mbox{\rm arbitrary} \, \alpha \\[1ex]
0, \qquad & r \rightarrow \infty.
\end{array}
\right.
\]

%%%%%%%%%%%%%%%%%%%%%%%%%%
\begin{figure}[ht]
\centering 
\hskip-0.5cm \epsfig{file=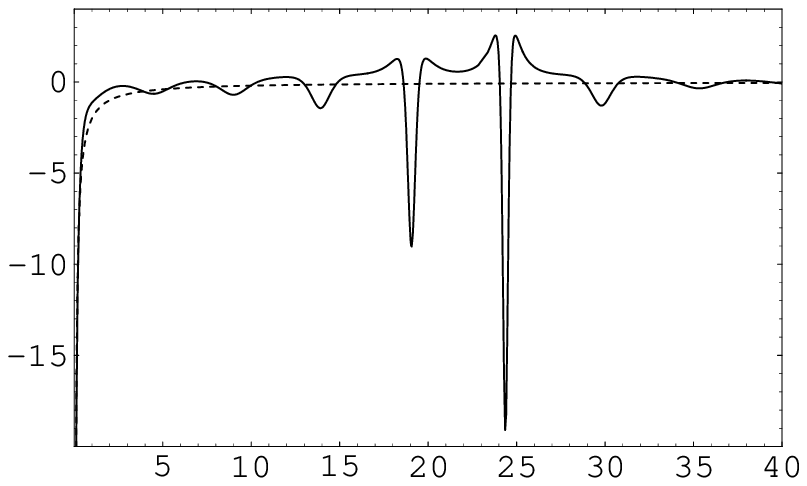, height=4cm}
\hskip1cm \epsfig{file=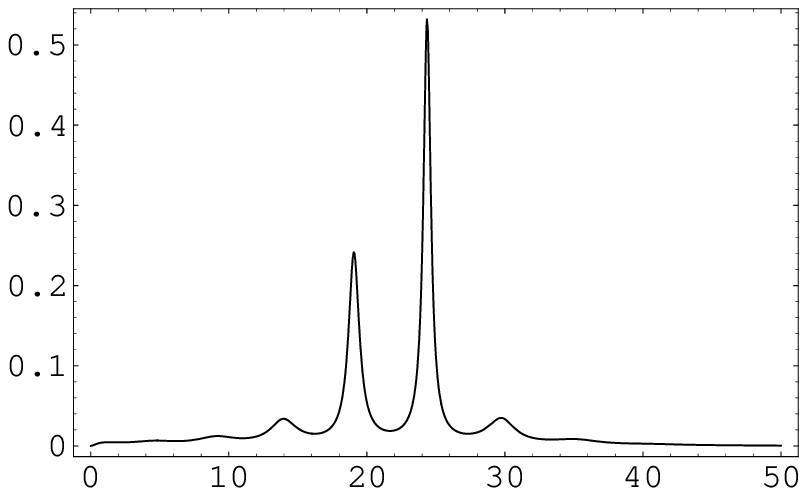, height=4cm}
\centerline{(a) \hskip6.7cm (b)}
\caption{\footnotesize {\bf (a)} The potential $V_{\ell - 1}(r)$ (dashed
curve) and the real part of $\mathsf{V}(r;\zeta)$, for $\ell =1$,
$\epsilon=-(0.1 + i \, 0.5)^2$, $\alpha=1$, and $\zeta = 0.5 + i \, 0.5$.
{\bf (b)} The corresponding unnormalized $\vert \Psi_{\epsilon}(r)
\vert^2$ for the same values of the parameters}
\end{figure} 
%%%%%%%%%%%%%%%%%%%%%%%%%

\noindent
The case ({\bf D}) of Appendix shows that, for $(\alpha, \zeta) \in
{\mathbb C}_0$ and $\theta_{\ell}(k) \neq 0$, the function
$\Psi_{\epsilon}(r)$ behaves as
\be
\Psi_{\epsilon}(r) \sim \left\{
\begin{array}{ll}
\frac{(2 k)^{2 \ell +1}}{\zeta} \, \frac{\Gamma(\ell +1 -1/k)}{\Gamma(2
\ell +1)} \, r^{\ell} & r \sim 0\\[2ex]
\frac{(2 k)^{\ell +1 +1/k}}{\alpha} \, \frac{\Gamma(\ell +1
-1/k)}{\Gamma(2 \ell + 2)} \, r^{1/k} \, e^{-kr} & r \rightarrow \infty,
\, k_1 >0 \\[2ex]
\frac{(2 k)^{\ell +1 -1/k}}{\theta_{\ell(k)}} \, r^{-1/k} \, e^{kr} &
r \rightarrow \infty, \, k_1 <0
\end{array}
\right.
\label{missing2}
\ee
which clearly belongs to $L^2({\mathbb R}^+, 4 \pi)$. Figure~2b depicts
the global behaviour of $\vert \Psi_{\epsilon} \vert^2$. Therefore, the
discrete spectrum $\sigma_d (\mathsf{H}_{\ell -1})$ of $\mathsf{H}(\zeta
\neq 0)$ is given by $\sigma_d(\mathsf{H}_{\ell -1}) =
\sigma_d(H_{\ell})\, \cup \, \{ \epsilon \}$, provided that $(\alpha,
\zeta) \in {\mathbb C}_0 , \, \theta_{\ell}(k) \neq 0$.

\noindent
Finally, the case ({\bf C}) of Appendix, gives another solution in the
same class.

%%%%%%%%%%%%%%%%%%%%%%%%%%%%%%%%%%%%%%%%%
\section{The new real hydrogen-like potentials}

We shall now extend the analysis of the previous section by intertwining
$H_{\ell}$ with a new (to be determined) Hamiltonian $\widetilde H$ as
follows
\be
\widetilde H \, \widetilde A = \widetilde A \, H_{\ell}
\label{intertwin2}
\ee
where the differential operator $\widetilde A$ is of second order
\be
\widetilde A := \frac{d^2}{dr^2} + \eta(r) \, \frac{d}{dr} + \gamma(r)
\label{seconda}
\ee
and $\widetilde H$ reads
\be
\widetilde H := - \frac{d^2}{dr^2} + \widetilde V(r).
\label{secondh}
\ee
The operators (\ref{seconda}--\ref{secondh}) depend implicitly on the
label $\ell$. A straightforward calculation allows to express the
functions $\eta$ and $\gamma$ of (\ref{seconda}) in terms of the
auto--B\"acklund transformation of the solutions of (\ref{riccati1}) for
$\epsilon_a$ and $\epsilon_b$ \cite{Ros98}:
\begin{eqnarray}
&& \eta(r) = - \left( \frac{\epsilon_a - \epsilon_b}{\beta_a(r)
-\beta_b(r)} \right), \qquad \epsilon_a \neq \epsilon_b,
\label{eta}\\[1ex]
&& \gamma(r) = \beta_b'(r) - \beta_b^2(r) + \eta(r) \, \beta_b(r), \qquad
\epsilon_a \neq \epsilon_b
\end{eqnarray}
Thus, the second order intertwining operator $\widetilde A$ in
(\ref{seconda}) is expressed by two different solutions of the first order
case. Moreover, it factorizes as $\widetilde A = a_2 a_1$, where
\be
a_1 \equiv \frac{d}{dr} + \beta_a = B; \qquad a_2 \equiv \frac{d}{dr} +
\eta - \beta_a = -A + \eta.
\label{first}
\ee
Thereby, it is easy to rewrite $\widetilde A$ as
\be
\widetilde A = (-A + \eta) B = -H_{\ell} + \epsilon + \eta B
\label{afactor}
\ee
and potential (\ref{secondh}) is obtained through
\be
\widetilde V (r) = V_{\ell}(r) + 2\, \eta'(r)
\label{system1}
\ee
In order to get a real $\eta(r)$ in (\ref{eta}, \ref{system1}), we
consider the solution $\beta_a(r)$ of (\ref{riccati1}) for $\epsilon_a \in
\mathbb{C}$ as given and, by taking $\epsilon_b = \bar \epsilon_a$ and
$\beta_b(r) = \bar \beta_a(r)$ in (\ref{eta}), one finds
\be
\eta(r) = - \frac{{\rm Im}(\epsilon_a)}{{\rm Im}(\beta_a)} \equiv
-\frac{\epsilon_2}{\beta_2(r)} = - \frac{d}{dr} \ln \omega(r)
\label{etareal}
\ee
where $\omega$ is defined in (\ref{w}) of Appendix and the labels $a$
and $b$ have been dropped from $\epsilon_2$ and $\beta_2$. Henceforth,
potential (\ref{system1}) is real:
\be
\widetilde V(r) 
= V_{\ell}(r) - 2 \, \frac{d^2}{dr^2} \, \ln \omega(r).
\label{potreal}
\ee
We are looking for potentials $\widetilde V(r)$ defined in the same
initial domain ${\cal D}_V=[0, \infty)$ (the situation when the initial
domain is changed requires a different treatment, see e.g. M\'arquez {\it
et. al.\/} in \cite{Mar98}.) According to the proposition of Appendix,
$\omega$ has at most one isolated zero in ${\cal D}_V$. By choosing a
proper $u$, the function $\omega$ can be so constructed that its isolated
zero coincides with one of the edges of ${\cal D}_V$ (compare
\cite{Fer02}):
\begin{eqnarray}
\lim\limits_{r \rightarrow 0} u(r)=0 \quad \Rightarrow \quad
\lim\limits_{r \rightarrow 0} \omega(r) =0
\label{limit1}\\[2.5ex]
\lim\limits_{r \rightarrow \infty} u(r)=0 \quad \Rightarrow \quad
\lim\limits_{r \rightarrow \infty} \omega(r) =0
\label{limit2}
\end{eqnarray}
Let us examine the consequences. First, condition (\ref{limit1}) is
satisfied if the $u$ in (\ref{u}) is chosen with $\alpha \neq 0$ and
$\zeta =0$, so that the potential (\ref{potreal}) behaves as shown on
Figure~3, {\it i.e.\/}:
\be
\widetilde V(r; \alpha \neq 0; \zeta=0) \sim \left\{
\begin{array}{ll}
V_{\ell +2}(r) & \mbox{\rm for}\, \, r \sim 0, \, \, \mbox{\rm arbitrary }
k_1\\[1ex]
0 & r \rightarrow \infty, \, \,  \mbox{\rm arbitrary } k_1
\end{array}
\right.
\label{potalpha}
\ee

%%%%%%%%%%%%%%%%%%%%%%%%%%
\begin{figure}[ht]
\centering 
\hskip-0.5cm \epsfig{file=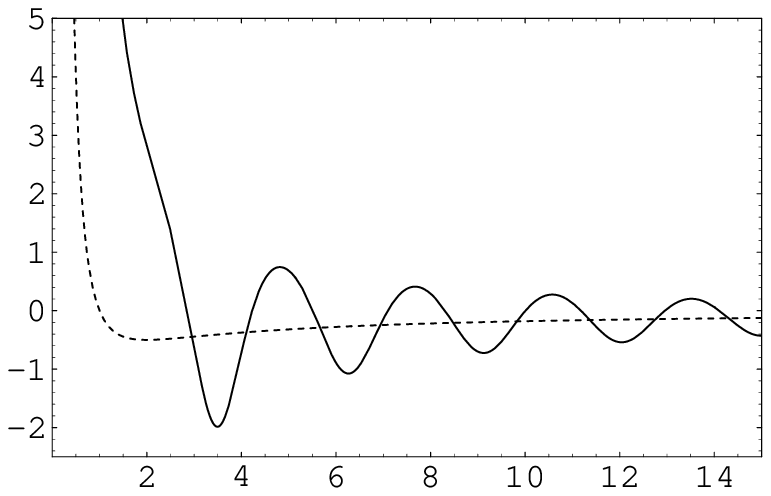, height=4cm}
\hskip1cm \epsfig{file=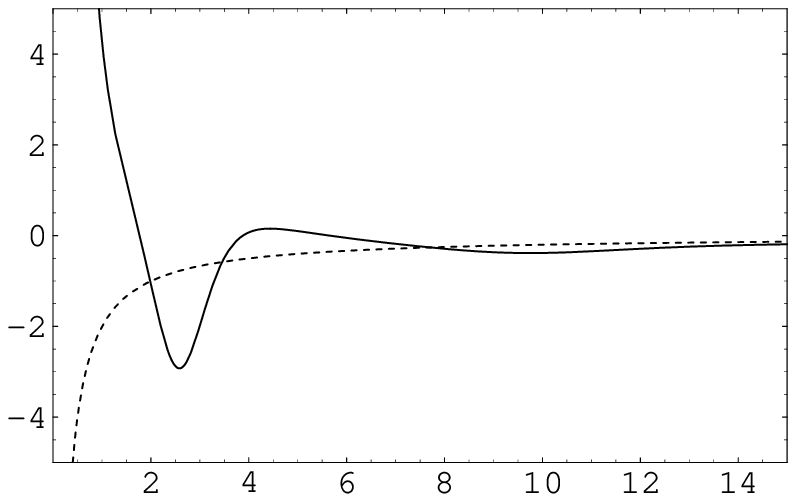, height=4cm}
\centerline{(a) \hskip6.7cm (b)}
\caption{\footnotesize 
The initial potential $V_{\ell}(r)$ (dashed curve) and its 2--susy partner
$\widetilde V(r)$ for $\alpha =1$, $\zeta=0$ and (a) $\ell=1$, $\epsilon =
-(0.01 + i)^2$ (b) $\ell=0$, $\epsilon = -(0.5 + i \, 0.1)^2$}
\end{figure} 
%%%%%%%%%%%%%%%%%%%%%%%%%
\medskip

\noindent
On the other hand, condition (\ref{limit2}) can be achieved for $\alpha
=0$, $\zeta \neq 0$ and $k_1>0$. In this case the equation (\ref{potreal})
leads to
\be
\widetilde V(r; \alpha= 0; \zeta \neq 0) \sim \left\{
\begin{array}{ll}
V_{\ell}(r) & \mbox{\rm for } r\sim 0,\\[1ex]
0 & r \rightarrow \infty.
\end{array}
\right.
\label{potzeta}
\ee
The real-valued potentials (\ref{potalpha}--\ref{potzeta}) resemble the
hydrogen one and they could, in principle, represent physical systems as
the Hamiltonians (\ref{secondh}) are self--adjoint. The next step is to
analize the new eigenvalue equation
\be
\widetilde H \widetilde \psi = \widetilde E \widetilde \psi
\label{eigen2}
\ee
whose solutions, by similar arguments as for the first order case, are
now obtained from the linear second order transformation (see equation
(\ref{intertwin2})):
\be
\widetilde \psi (r) \propto \widetilde A \, \psi_{n, \ell}(r) = (-E_n +
\epsilon) \, \psi_{n, \ell}(r) + \eta (r, \zeta) \, \Psi(r, \zeta)
\label{psidos}
\ee
where we have used (\ref{afactor}) and $\Psi(r, \zeta)$ is given by
(\ref{Psi2}). The corresponding boundary conditions at $r=0$ can be
obtained from those of $\psi_{n, \ell}(r)$ and $\Psi(r, \zeta)$.

It is clear that the first term at the r.h.s. of
(\ref{psidos}) is in $L^2({\mathbb R}^+, 4 \pi)$. The behaviour of the
related second term is found by observing that

1) If $\alpha \neq 0$ and $\zeta =0$ then $\eta$ diverges as $r^{-1}$ at
the origin while it is constant at $r=\infty$. Therefore the product
$\eta(r,0) \, \Psi(r,0)$ is zero at the edges of ${\cal D}_V$ and remains
finite in all ${\cal D}_V$.

2) If $\alpha = 0$ and $\zeta \neq 0$ then $\eta$ is a constant at both
edges of ${\cal D}_V$. Hence, $\eta(r, \zeta \neq 0)  \, \Psi(r,\zeta \neq
0)$ is again well behaved in all ${\cal D}_V$.

Thus, the eigenfunctions $\widetilde \psi$ of $\widetilde H$ given in
(\ref{psidos}) are in $L^2({\mathbb R}^+,
4 \pi)$ and $\widetilde H$ is an exactly solvable Hamiltonian with the
same spectrum as the hydrogen atom.

%%%%%%%%%%%%%%%%%%%%%%%%%%%%%%%%%%%%%%%%
\section{Summary and Discussion}

In this paper we have used a new type of factorization method to analyze a
set of non--hermitian susy partners of the radial part of the hydrogen
Hamiltonian. In order to generate the corresponding potentials we have
used Darboux transformations with complex factorization constants. In
contrast with {\it PT\/}--symmetry \cite{Ben98, Ben99b, Dor01} and
pseudo--supersymmetry \cite{Mos02}, the breaking of supersymmetry leads to
purely real spectra. However, while the pseudo--hermiticity and {\it
PT\/}--symmetry breaking involve pairs of conjugate complex eigenvalues,
the unbroken supersymmetry (for which the non--hermitian 1--susy partners
are not strictly isospectral) involves just a single `complex energy'. To
be more precise, in order to add two extra eigenvalues (real or complex)
to a given spectrum one applies either twice the 1--susy procedure or a
single 2--susy transformation (both can be made equivalent for the case we
are dealing with.) Now, if the two new energies form a complex conjugate
pair and if the functions $\eta(r)$ and $\gamma(r)$ of the non--singular
intertwining operator (\ref{seconda}) are real, then the final Hamiltonian
becomes self-adjoint and does not admit any complex eigenvalue, just as we
have shown in Section~4.

The problem of finding normalizable solutions belonging to complex
eigenvalues for non--hermitian Schr\"odinger equations has been solved
previously by numerical techniques in \cite{Bay96} and analized inside a
Lie--algebraic framework in \cite{Bag02}. In this paper we performed an
analytical study and we expect that our results complement the numerical
ones. Similarly as in other non--hermitian cases discussed in the
literature, the interpretation of the `complex energies' is an open
problem, though notice possible applications to the absorptive
(dissipative) systems \cite{last}. 

In general, we have shown that, the extension of the SUSY treatment to
include complex factorization constants leads to results which are out of
the scope of the {\it PT\/}--symmetry and pseudo--hermiticity. Indeed, the
reality of the spectrum of the Hamiltonians ${\mathsf H} (\zeta)$ in
Section~3 depends on the parameter $\zeta$, as it has been established in
Subsections~3.1 and 3.2. As the non--hermitian Hamiltonians ${\mathsf H}
(\zeta \neq 0)$ in Subsection~3.2 do not satisfy the theorems by
Mostafazadeh \cite{Mos02}, they are not pseudo--hermitian. On the other
hand, it is not yet clear wheter the Hamiltonians ${\mathsf H} (\zeta=0)$
of Subsection~3.1 could be pseudo--hermitian or not (though a primary
impression can be depicted by noticing that the non--hermiticity of
${\mathsf H} (\zeta)$ depends not on $\zeta$ but on the non--trivial
imaginary part of the factorization constant $\epsilon$.) Work in this
direction is in progress and will be published elsewhere.

\vskip0.5cm
\noindent
{\bf Acknowledgements.} The support of CONACyT (M\'exico), project
40888-F, is acknowledged.

\bigskip

%%%%%%%%%%%%%%%%%%%%%%%%%%%%%%%%%%%%%%%%%%%%%%%%%%%%%%%%%%%%%%
%%%%%       APPENDIX'S MACROS   %%%%%%%%%%%%%%%%%%%%%%%%%%%%%%
%%%%%%%%%%%%%%%%%%%%%%%%%%%%%%%%%%%%%%%%%%%%%%%%%%%%%%%%%%%%%%
\renewcommand{\thesection}{Appendix}
\setcounter{section}{0}
\renewcommand{\theequation}{\Alph{section}.\arabic{equation}}
\setcounter{equation}{0}
%%%%%%%%%%%%%%%%%%%%%%%%%%%%%%%%%%%%%%%%%%%%%%%%%%%%%%%%%%%%%%

\section{}

The global behavour of the eigenfunctions $u(r)$ of equation (\ref{u}) can
be described in terms of $\alpha$, $\zeta$ and the sign of $k_1$:

{\bf A}) If $\alpha \neq 0$ and $\zeta =0$ then $u(r)$ is zero at the
origin $r=0$ while it diverges at $r= \infty$.

{\bf B}) If $\alpha = 0$, $\zeta \neq 0$ and $k_1>0$, then $u(r)$
diverges at the origin and tends to zero for $r \rightarrow \infty$.

{\bf C}) If $\alpha = 0$, $\zeta \neq 0$ and $k_1<0$, then $u(r)$
diverges at $r=0$ and  $r= \infty$.

{\bf D}) Let ${\mathbb C}_0 \subset {\mathbb C} \times {\mathbb C}$ be the
subset of complex pairs $\alpha \neq 0$, $\zeta \neq 0$, such that
\[
\zeta \neq  - \alpha \, \frac{{}_1F_1(\ell +1 -1/k, 2 \ell +2, 2 k
r_0)}{U(\ell +1 -1/k, 2 \ell +2, 2 k r_0)}, \qquad \forall r_0 \in (0,
\infty).
\]
\indent
If
\be
\theta_{\ell} (k):= \zeta + \alpha \, \frac{\Gamma (2 \ell +
2)}{\Gamma(\ell +1 + 1/k)} \, e^{ \pm i (\ell + 1 - 1/k) \pi}
\label{gamma}
\ee
is different from zero for $(\alpha, \zeta) \in {\mathbb C}_0$ , then
$u(r)$ is free of zeros in all ${\cal D}_V$ and diverges at $r=0$ and
$r=\infty$.

{\bf E}) If $(\alpha, \zeta) \in {\mathbb C}_0$, $k_1<0$ and
$\theta_{\ell}(k) =0$, then $u(r)$ diverges at the origin while $\lim_{r
\rightarrow \infty} u(r)=0$ (coinciding indeed with {\bf B}).

\noindent
The presence of zeros in these functions has been studied by means of the
following

%%%%%%%%%%%%%%%%%%%%%%%%%%%%%%%%%%%%%%%%%%%%
\begin{itemize}
\item[]

{\it Proposition\/}: Let $u(r) \in C^1(D_v)$ be solution of the
Schr\"odinger equation $u''(r) = [v(r) - \epsilon] \,u(r)$, where $v(r)$
is a real-valued potential with domain ${\cal D}_v$ and $\epsilon \in
{\mathbb C}$. Assume that ${\cal D}_v$ is a simply connected region of
${\mathbb R}$. If ${\rm Im} \, (\epsilon) \neq 0$, then the complex-valued
function $u(r)$ admits at most one isolated zero in ${\cal D}_v$.

{\it Proof\/}: Let
\be
\omega(r) := \frac{W(u, \bar u)}{2\, i \, {\rm Im} \, (\epsilon)}
\label{w}
\ee
where the bar denotes complex conjugation and $W(\cdot, \cdot)$
corresponds to the Wronskian of the involved functions. Clearly $\omega$
is continuous on ${\cal D}_v$ and $\omega'(r) = \vert u (r) \vert^2 \geq
0$ $\forall \, r \in {\cal D}_v$, so $\omega(r)$ is always non--decreasing
and can have either only one isolated zero or an entire interval of zero
points in ${\cal D}_v$. As every zero of $u(r)$ is, necessarily, a zero of
$\omega(r)$ then $u(r)$ admits at most one isolated zero there.
$\diamond$

\end{itemize}
%%%%%%%%%%%%%%%%%%%%%%%%%%%%%%%%%%%%%%%%%%%%%%%%%%%%%%%%%%%%%

\noindent
The real function $\omega(r)$ in (\ref{w}) plays a relevant role in the
2--susy approach of Section~4. A convenient expression for $\beta(r)$ in
terms of $\omega$ is given by
\be
\beta(r) = \beta_1(r) + i\, \beta_2(r) \equiv -\frac12 \frac{d}{dr} \ln
\omega'(r) + i\, \epsilon_2 \left( \frac{d}{dr} \ln \omega(r)
\right)^{-1}.
\label{betas}
\ee

%%%%%%%%%%%%%%%%%%%%%%%%%%%%%%%%%%%%%

%%%%%%%%%%%%%%%%%%%%%%%%%%%%%%%%%%%%%

\end{document}